\documentclass[prb,twocolumn,showpacs,amsmath,amssymb,floatfix]{revtex4}
\usepackage{color,graphics,epsfig,rotating}
\usepackage{graphicx}
\usepackage{dcolumn}
\usepackage{bm}
\usepackage{subfigure}

\begin{document}
\title{First-principles study of the electronic structure and magnetism of
  CaIrO$_3$}

\author{Alaska Subedi}
\affiliation{Max Planck Institute for Solid State Research,
  Heisenbergstrasse 1, D-70569 Stuttgart, Germany}

\begin{abstract}
  I study the electronic structure and magnetism of postperovskite
  CaIrO$_3$ using first-principles calculations. The density
  functional calculations within the local density approximation
  without the combined effect of spin-orbit coupling and on-site
  Coulomb repulsion show the system to be metallic, which is in
  disagreement with the recent experimental evidences that show
  CaIrO$_3$ to be an antiferromagnetic Mott insulator in the
  $J_\textrm{eff}$ = 1/2 state. However, when spin-orbit coupling is
  taken into account, the Ir $t_{2g}$ bands split into fully filled
  $J_\textrm{eff}$ = 3/2 bands and half-filled $J_\textrm{eff}$ = 1/2
  bands.  I find that spin-orbit coupling along with a modest on-site
  Coulomb repulsion opens a gap leading to a Mott insulating state.
  The ordering is antiferromagnetic along the $c$ axis with total
  moments aligned antiparallel along the $c$ axis and canted along the
  $b$ axis.
\end{abstract}


\maketitle

\section{Introduction}

Transition-metal oxides (TMOs) in perovskite and related structures
exhibit myriad interesting properties. These include unconventional
superconductivity in cuprates,\cite{bedn86} colossal magnetoresistance
in manganites,\cite{helm93} and ferroelectricity in
Pb(Zr,Ti)O$_3$.\cite{jaff71}
In these materials, the transition-metal ion is situated inside an
oxygen octahedral cage, which may be arranged in a corner- or
edge-shared manner.
The interesting properties of TMOs arise because of the competition
between the crystal-field splitting (which arises because of covalency
between transition metal $d$ and oxygen $p$ states), on-site Coulomb
repulsion $U$, Hund's coupling, spin-orbit (SO) coupling due to
orbital degeneracy (leading to unquenched angular moment in the ground
state), and different spin-exchange pathways.

In this regard, the discovery of a spin-orbital Mott state in
Sr$_2$IrO$_4$ by Kim \textit{et al.}\cite{kim08,kim09} is significant
because it enables us to study the case where spin-orbit coupling and
its interplay with the Coulomb repulsion is an important ingredient in
determining the electronic and magnetic properties of the system.
Sr$_2$IrO$_4$ exists in a layered perovskite structure. The Ir$^{4+}$
(5$d^5$) ions are situated inside corner-shared O octahedral cages,
which are themselves arranged in a square lattice in the $ab$ plane.
As there are an odd number of electrons per formula unit, one might
expect this material to be a metal in the band picture. However,
Sr$_2$IrO$_4$ is experimentally a canted antiferromagnetic
insulator.\cite{kim09}
As explained lucidly by Kim \textit{et al.}, a state with an effective
total angular momentum $J_\textrm{eff} = 1/2 $ that has a complex wave
function is realized in Sr$_2$IrO$_4$, which arises due to the
combined influence of strong spin-orbit coupling and moderate on-site
Coulomb repulsion.\cite{kim08} A similar conclusion on Sr$_2$IrO$_4$
has been reached through the study of a three-orbital Hubbard model
with spin-orbit coupling\cite{wata10} and combined density functional
theory and dynamical mean field theory (LDA+DMFT)
calculations,\cite{mart11} although Arita \textit{et al.}\ suggest
that Sr$_2$IrO$_4$ is a Slater insulator based on their LDA+DMFT
study.\cite{arit11} Interestingly, it was also found that spin-orbit
coupling plays an important role in the electronic properties even for
a 4$d^5$ system such as Sr$_2$RhO$_4$.\cite{have08,liu08}

\begin{figure} 
  \includegraphics[width=0.9\columnwidth]{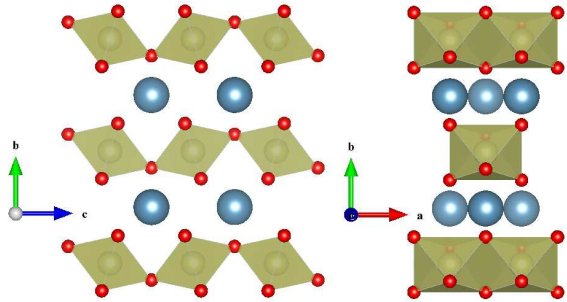}
  \caption{(Color online) Crystal structure of CaIrO$_3$. The large
    (cyan) balls are Ca, small (red) balls are O, and the Ir atoms
    reside inside the (brown) octahedra.}
  \label{fig:cio-struct}
\end{figure}

It has been suggested that systems in the $J_\textrm{eff}$ = 1/2
state, depending on bond geometry, lead to interesting varieties of
low-energy Hamiltonians, including the isotropic Heisenberg model and
the highly anisotropic quantum compass or Kitaev models relevant for
quantum computing.\cite{jack09}
Therefore, it is important to investigate materials that exhibit the
$J_\textrm{eff} = 1/2$ state in different structures in order to study
the effect of different local environments and spin-exchange pathways.
Recently, Ohgushi \textit{et al.}\ reported resonant x-ray diffraction
study of CaIrO$_3$ that indicates this material also exhibits a Mott
insulating $J_\textrm{eff} = 1/2$ state.\cite{ohgu11}
CaIrO$_3$ exists in the postperovskite structure with space group
$Cmcm$ as shown in Fig.~\ref{fig:cio-struct}. The Ir$^{4+}$ (5$d^5$)
ions are situated inside the O octahedra, but these octahedra share an
edge along the $c$ axis, unlike the case of Sr$_2$IrO$_4$.
Thus, CaIrO$_3$ is another ideal material to investigate the interplay
between spin-orbit coupling and on-site Coulomb repulsion that may
help in understanding the unique properties that might be exhibited by
$J_\textrm{eff} = 1/2$ systems.

Tsuchiya \textit{et al.}\ have reported first-principles density
functional calculations that show this material is a metal within the
local density approximation.\cite{tsuc07} This is contrary to the
experimental evidence that shows this material is a Mott insulator
that undergoes an antiferromagnetic transition at $T_\textrm{N}$ = 115
K.\cite{ohgu06}
A recent resonant x-ray diffraction study shows that the ordering is of
stripe-type antiferromagnetism along the $c$ axis, with total moments
aligning parallel along the $a$ axis and antiparallel along the
$c$ axis.\cite{ohgu11} The inverse susceptibility 1/$\chi$ deviates
from the linear behavior at a temperature $\sim$350 K that is
considerably higher than $T_\mathrm{N}$ and a Curie-Weiss fit to
$\chi$ above 400 K gives a Curie-Weiss temperature of 3900
K.\cite{ohgu06} This indicates that the antiferromagnetic correlations
arise much before the antiferromagnetic transition, and magnetic
ordering is suppressed by low dimensionality or competing ordering
interactions.
Jang \textit{et al.} have studied the electronic structure of
meta-stable perovskite Ca$_{1-x}$Sr$_x$IrO$_3$ ($x$ = 0, 0.5, and 1)
thin films using transport measurements, optical spectroscopy, and
pseudopotential-based first-principles calculations.\cite{jang10} They
find that perovskite CaIrO$_3$ thin films are semimetallic and near
the metal-insulator transition. Their calculations with spin-orbit
coupling and on-site Coulomb repulsion $U$ found that the spin-orbit
coupling splits the Ir $t_{2g}$ states into $J_\textrm{eff}$ = 3/2 and
1/2 states, and $U$ = 2.0 eV further splits the $J_\textrm{eff}$ = 1/2
states, although the valence and conduction bands still touch the
Fermi level, resulting in a semimetallic state.

The experimental evidences that have so far been accumulated suggest
that calculations that include the effect of spin-orbit coupling and
on-site Coulomb repulsion would be helpful in clarifying the
electronic and magnetic properties of CaIrO$_3$.  In this paper, I
report the results of density functional calculations that show
CaIrO$_3$ is in a Mott insulating state that is induced by the
combined effect of spin-orbit coupling and on-site Coulomb repulsion.
This state arises out of spin-orbit split Ir $t_{2g}$ bands that get
separated into lower lying fully filled and higher lying half-filled
bands that have effective total angular momenta $J_\textrm{eff}$ =
3/2 and 1/2, respectively, in the strong spin-orbit coupling limit.
The half-filled $J_\textrm{eff}$ = 1/2 bands are narrow, so even a
modest on-site Coulomb repulsion induces a Mott insulating state that
is topologically different from the metallic state given by the local
density approximation, without taking into account the spin-orbit
coupling and on-site Coulomb repulsion.
This is a Mott insulating state in the sense that a single-particle
theory such as the density functional theory implemented using
Kohn-Sham formalism cannot explain the insulating state, and an
explicit treatment of on-site Coulomb repulsion is needed.
The Mott insulating state thus obtained is antiferromagnetically
ordered along the $c$ axis with total moments aligned antiparallel
along the $c$ axis and canted along the $b$ axis.

\section{Approach}
The purpose of this paper is to elucidate the role of spin-orbit
coupling and on-site Coulomb repulsion on the electronic and magnetic
properties of CaIrO$_3$ using density functional calculations. The
calculations were performed within the local density approximation
(LDA) using the general full-potential linearized augmented plane-wave
method as implemented in the ELK software package.\cite{elk} Muffin-tin
radii of 2.2, 2.0, and 1.6 a.u. for Ca, Ir, and O, respectively, were
used. A $8 \times 8 \times 8$ $k$-point grid was used to perform the
Brillouin zone integration, and the convergence of moments was checked
on a $10 \times 10 \times 10$ grid. The effect of spin-orbit coupling
was treated using a second-variational scheme, and the fully localized
limit\cite{fll} is used to take into account the double counting in
LDA+$U$ calculations. A value for the on-site Coulomb repulsion $U$ =
2.75 eV (which gives a band gap close to the experimental value) was
used unless otherwise mentioned.

I used the experimental lattice parameters $a$ = 3.145 \AA, $b$ =
9.855 \AA, and $c$ = 7.293 \AA,\cite{rodi65} but relaxed the atomic
positions. The calculated atomic positions Ca (0, 0.2498, 0.25), Ir
(0, 0, 0), O(1) (0.5, 0.4253, 0.25), and O(2) (0.5, 0.1230, 0.0485) agree
well with the experimental values Ca (0, 0.2498, 0.25), Ir (0, 0, 0),
O(1) (0.5, 0.4331, 0.25), and O(2) (0.5, 0.1296, 0.0553). The results
presented in this paper are for the relaxed atomic positions, but I
also performed calculations with experimental atomic positions and
came to the same physical conclusions.
There are two formula units per primitive unit cell in the $Cmcm$
structure. The Ir$^{4+}$ ions make a two-dimensional rectangular
lattice in the $ac$ plane (not shown) and the Ir-O layer is separated
by a layer of Ca atoms along the $b$ axis. For the relaxed atomic
positions, the O octahedra are tilted by an angle of $22^\circ$. The O
octahedra are slightly compressed along the corner-shared O direction
(from left to right in the left figure of Fig.~\ref{fig:cio-struct})
with a bond-length ratio of 0.97. The Ir-O distances along the
corner-shared $c$ axis and edge-shared $a$ axis are 1.97 and
2.02 \AA, respectively.

\begin{figure} 
  \includegraphics[width=0.95\columnwidth]{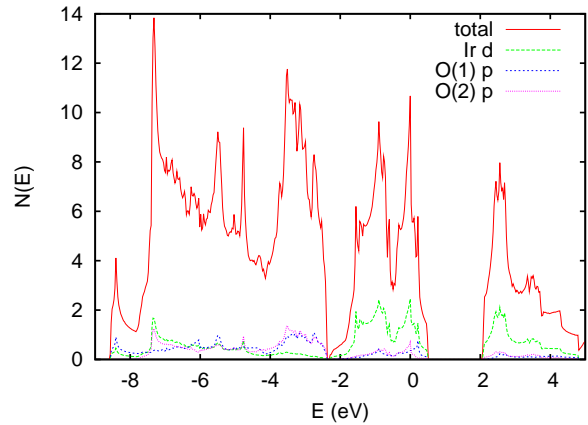}\\
  \caption{(Color online) Non-spin-polarized LDA DOS of CaIrO$_3$ (in
    states/eV). The projections are onto the respective muffin-tin
    spheres and are only indicative of the contribution to the total
    DOS. The Fermi energy is at 0 eV.}
  \label{fig:nsp-lda}
\end{figure}

\section{Results}
\begin{figure}
  \centering
  \includegraphics[width=\columnwidth]{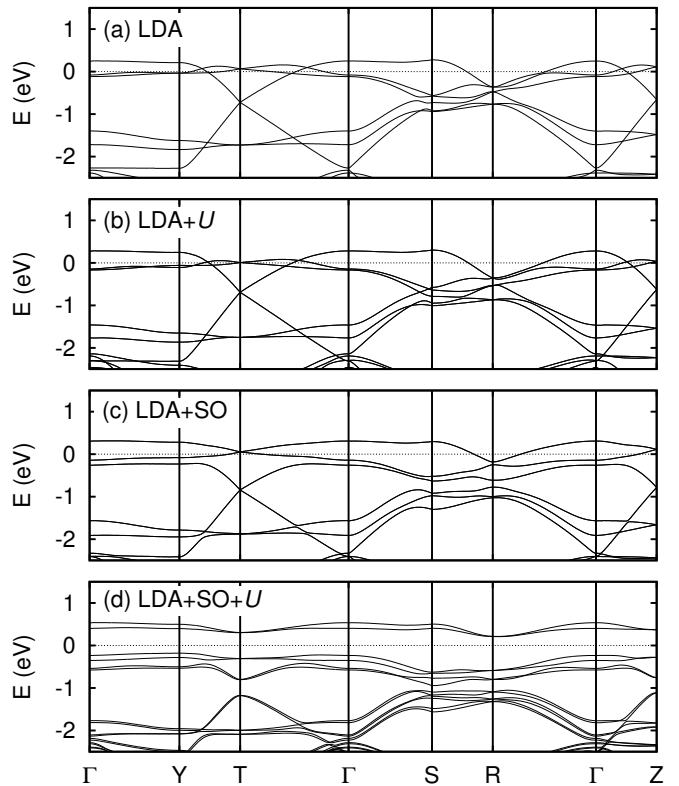}
  \caption{NSP LDA, LDA+$U$, LDA+SO and LDA+SO+$U$ band structures,
    respectively from top to bottom, of CaIrO$_3$. The band structures
    are plotted along the path $\Gamma$ (0,0,0) $\rightarrow$ Y
    (0,$\frac{1}{2}$,0) $\rightarrow$ T (0, $\frac{1}{2}$,
    $\frac{1}{2}$) $\rightarrow$ $\Gamma$ (0,0,0) $\rightarrow$ S
    ($\frac{1}{4}$, $\frac{1}{4}$, 0) $\rightarrow$ R ($\frac{1}{4}$,
    $\frac{1}{4}$, $\frac{1}{2}$) $\rightarrow$ $\Gamma$ (0,0,0)
    $\rightarrow$ Z (0,0,$\frac{1}{2}$). Here, $U$ = 2.75 eV is
    used. The bands are exchange split only for the case of
    LDA+SO+$U$.}
  \label{fig:bnd-comb}
\end{figure}

Let us first consider the non-spin-polarized (NSP) LDA
calculations. Even though they are inadequate to describe the
ground-state properties, these calculations give a decent description
of the band structure of CaIrO$_3$ that provide a playground for the
interplay between SO coupling and $U$.
The electron density of states (DOS) and band structure for this case
are shown in Figs.~\ref{fig:nsp-lda} and ~\ref{fig:bnd-comb}(a),
respectively.
Most of the bands along the $b$ direction ($\Gamma$-$Y$) have low
dispersion, which suggests that the physics related to two dimensionality
might be relevant in this system.
There are 24 bands between $-$9.0 and 1.0 eV. The 18 bands between
$-$9.0 and $-$2.3 eV (not shown) have a dominant O $p$ character and
thus derive from the $p$ orbitals of the six O atoms in the unit
cell. These bands also show Ir $d$ character, which implies
significant covalency between the O $p$ and Ir $d$ states as the
unoccupied Ir $d$ states above Fermi level also contain some admixture
of O $p$ states. There is a very small gap of $\sim$0.05 eV at $-$2.36
eV, beyond which lie six bands with a dominant Ir $d$ character. These
are the Ir $t_{2g}$ states that are formally antibonding, and these
bands correspondingly show some O $p$ contribution.
A gap of $\sim$1.5 eV separates these $t_{2g}$ states from a group of
four bands that have a mostly Ir $d$ character, which are the Ir $e_g$
states. The Ir $e_g$ also have some O $p$ character due to Ir $d$--O
$p$ covalency.
The Ir 5$d$ states are quite delocalized and the edge-sharing
compressed IrO$_6$ octahedra are rotated by $22^\circ$, and this leads
to some hybridization the between Ir $t_{2g}$ and $e_g$ levels.
The Ca and Ir $s$ states are high above the Fermi level, and within an
ionic limit the electronic structure is consistent with the ionic
states Ir$^{4+}$ and O$^{2-}$, although there is significant deviation
from this because of Ir $d$--O $p$ hybridization.
The two Ir$^{4+}$ ions nominally have five electrons each in their $d$
orbitals. As a result, the six Ir $t_{2g}$ bands are not fully filled,
and the system is a metal within LDA with a $t_{2g}$ hole on each
Ir$^{4+}$ ion.

The metallic state obtained within LDA due to incomplete filling of Ir
$t_{2g}$ states is contrary to the experimental evidence that
indicates CaIrO$_3$ is a Mott insulator. This suggests that spin-orbit
coupling and/or on-site Coulomb repulsion play crucial roles in the
electronic and magnetic properties of CaIrO$_3$.
Let us now consider the effect of SO coupling and $U$ on the
electronic structure of CaIrO$_3$. The NSP LDA, LDA+$U$, LDA+SO,
and LDA+SO+$U$ (with $U$ = 2.75 eV) Ir $t_{2g}$ bands are shown in
Fig.~\ref{fig:bnd-comb}.
Let us first note that a value for $U$ of 2.75 eV without the
spin-orbit coupling has very little effect on the band structure (I
did the calculation with $U$ up to 5 eV without getting an insulating
state). This is not surprising as the Ir $t_{2g}$ manifold is spread
over a bandwidth of $\sim$2.8 eV, and it would require a substantially
larger $U$ to open up a gap.
However, turning on spin-orbit coupling makes a significant difference
in the electronic structure. The system is non-magnetic, so the bands
are spin degenerate as they are not exchange split. However, the
spin-oribt coupling splits the manifold of six spin-degenerate Ir
$t_{2g}$ bands into a lower-lying group of four and a higher-lying
group of two spin-degenerate bands.
In the limit of strong spin-orbit coupling, the lower and higher sets
of bands within the $t_{2g}$ manifold would correspond to effective
total angular momenta $J_\textrm{eff}$ of 3/2 and 1/2,
respectively.
This is similar to the case of Sr$_2$IrO$_4$ where the spin-orbit
coupling splits the Ir $t_{2g}$ bands into a lower lying quartet of
$J_\textrm{eff}$ = 3/2 and a higher lying doublet of $J_\textrm{eff}$
= 1/2 bands.\cite{kim08}
In the case of CaIrO$_3$, the $J_\textrm{eff}$ = 1/2 bands are narrow
with a width of $\sim$1 eV. These are separated from the
$J_\textrm{eff}$ = 3/2 bands by $\sim$0.15 eV, although the gap is
indirect. The ten Ir $d$ electrons in the unit cell completely fill the
$J_\textrm{eff}$ = 3/2 bands, while the $J_\textrm{eff}$ = 1/2 bands
are only half filled.  As a result, the system is still metallic.

\begin{table*}[!htp]
  \caption{\label{tab:expect} The $\langle \vec{L} \rangle$ and $\langle \vec{S}
    \rangle$ expectation values computed over Ir muffin-tin
    spheres and the band gap $E_\textrm{gap}$ (eV) for some values of
    on-site Coulomb repulsion $U$ (eV) and Hund's coupling $J$ (eV). 
    The moments are in units of Bohr magneton.}
  \begin{ruledtabular}
    \begin{tabular}{lccl}
      Site & $\langle \vec{L} \rangle$ & $\langle \vec{S} \rangle$
      & \\
      \hline
      Ir(1) & $(0.00, 0.06, -0.28)$ &  $(0.00, 0.03, -0.19)$ & $U
      = 2.75$, $J = 0.0$ \\
      Ir(1$^\prime$) & $(0.00, 0.06, 0.28)$ & $(0.00, 0.03,
      0.19)$ & $E_\textrm{gap} = 0.33$ \\
      \hline
      Ir(1) & $(0.00, 0.06, -0.27)$ &  $(0.00, 0.03, -0.18)$ & $U
      = 2.75$, $J = 0.3$ \\
      Ir(1$^\prime$) & $(0.00, 0.06, 0.27)$ & $(0.00, 0.03,
      0.18)$ & $E_\textrm{gap} = 0.30$ \\
      \hline
      Ir(1) & $(0.00, 0.04, -0.18)$ &  $(0.00, 0.02, -0.13)$ & $U
      = 2$, $J = 0.0$ \\
      Ir(1$^\prime$) & $(0.00, 0.04, 0.18)$ & $(0.00, 0.02,
      0.13)$ & $E_\textrm{gap} = 0.03$ \\
    \end{tabular}
  \end{ruledtabular}
\end{table*}

Even though on-site Coulomb repulsion $U$ and spin-orbit coupling
acting alone do not make the system an insulator, it is likely that
their combined effect can induce a Mott insulating state by splitting
the narrow $J_\textrm{eff}$ = 1/2 bands. The LDA+SO+$U$ calculations
with $U$ = 2.75 eV reveal that this scenario is realized in
CaIrO$_3$. As shown in Fig.~\ref{fig:bnd-comb}(d), the $U$ in the presence
of SO coupling makes only minor modifications to the $J_\textrm{eff}$
= 3/2 bands.
The $J_\textrm{eff}$ = 3/2 bands get exchange split and a degeneracy
at the point $T$ $(0,0.5,0.5)$ is lifted, but otherwise the bandwidth
and topology of the bands do not change substantially.
However, the half-filled $J_\textrm{eff}$ = 1/2 bands, in addition to
being exchange split by $\sim$0.1 eV, are split into the upper (UHB)
and lower (LHB) Hubbard bands, yielding a Mott insulating state. The
fully occupied $J_\textrm{eff}$ = 1/2 LHB has a small bandwidth of
$\sim$0.5 eV and is separated from the unoccupied UHB by a gap of
$\sim$0.33 eV for $U$ = 2.75 eV (the gap is $\sim$0.03 eV for $U$ = 2
eV). This agrees well with a band gap of 0.34 eV obtained
experimentally.\cite{ohgu06,foot:gap}
I also performed calculations with $U$ = 1.0, 1.5, 2.0, and 2.5
eV. The system is metallic within LDA+SO+$U$ for $U$ up to 1.5 eV, but
it becomes an insulator by $U$ = 2.0 eV. To see the effect of the Hund
coupling $J$, I did LDA+SO+$U$ calculations with $U$ = 2.75 eV and $J$
= 0.1, 0.2, and 0.3 eV. I find that these values of Hund coupling $J$
do not change the qualitative picture---the Ir $t_{2g}$ levels are
still split into $J_\textrm{eff}$ = 3/2 and 1/2 bands and the
$J_\textrm{eff}$ = 1/2 bands are further split into fully occupied LHB
and unoccupied UHB. The inclusion of Hund coupling mainly reduces the
band gap (for $J$ = 0.3 eV, the band gap is 0.30 eV) and the magnetic
moment.

The LDA+SO+$U$ calculations give an antiferromagnetic ground state for
CaIrO$_3$ along the $c$ axis with total moments aligning antiparallel
along the $c$ axis. The orbital and spin moments are parallel to each
other along the $c$ axis and ferromagnetically canted along the
$b$ axis. The angular and spin expectation values computed over the
two Ir muffin-tin spheres Ir(1) (0.0, 0.0, 0.0) and Ir(1$^\prime$)
(0.0, 0.0, 0.5) for different $U$ and $J$ values are given in Table
\ref{tab:expect}. For $U$ = 2.75 eV, the total moment is 0.67
$\mu_B$/Ir with an orbital moment of 0.29 $\mu_B$ and a spin moment of
0.38 $\mu_B$ ($ = 2 | \langle \vec{S} \rangle |$). The canting angle
is 10$^\circ$, approximately half the octahedral tilting angle of 22$^\circ$
and twice the value of 4$^\circ$ reported in
Ref.~\onlinecite{ohgu11}. The calculated values differ considerably
from what is expected for the ideal $J_\textrm{eff}$ = 1/2 state. In
the ionic limit, one expects an orbital moment of 0.67 $\mu_B$ and a spin
moment of 0.33 $\mu_B$ for a $J_\textrm{eff}$ = 1/2 state.\cite{kim08}
In contrast, I obtain an orbital moment that is lower than the spin
moment.  The reason for this deviation from the $J_\textrm{eff}$ = 1/2
may be the compression and tilting of the IrO$_6$ octahedra, in
addition to the covalency between Ir $d$ and O $p$ states. The
compression of the IrO$_6$ octahedra will quench the orbital moment as
the degeneracy between $t_{2g}$ states are lifted. Also, the tilting
of the IrO$_6$ octahedra causes the $e_g$ bands to get mixed with the
$t_{2g}$ states. These two effects should reduce the orbital moment
but might enhance the spin contribution. It is interesting to note
that Sr$_2$IrO$_4$ also has distortion of the IrO$_6$ octahedra with
a bond-length ratio of 1.04 and a tilting angle of
11$^\circ$,\cite{craw94} and it has a calculated orbital moment of 0.26
$\mu_B$ and spin moment of 0.10 $\mu_B$.\cite{kim08} As the tilting
angle in CaIrO$_3$ is twice that of Sr$_2$IrO$_4$, it might be
reasonable to expect that CaIrO$_3$ deviates further from the ideal
$J_\textrm{eff}$ = 1/2 state due to the mixing of the $e_g$ states.

\section{Conclusions}
In summary, the electronic structure and magnetic properties of
CaIrO$_3$ has been studied using first-principles calculations. The
system is metallic within the LDA because the Ir $t_{2g}$ states are
incompletely filled. Modest values of on-site Coulomb repulsion alone
have very little effect on the LDA electronic structure as the Ir
$t_{2g}$ states have a broad bandwidth. The introduction of spin-orbit
coupling splits the Ir $t_{2g}$ states into fully filled
$J_\textrm{eff}$ = 3/2 bands and half-filled $J_\textrm{eff}$ = 1/2
bands. The half-filled bands have a small bandwidth of $\sim$1.0 eV,
which is split into fully filled lower and unfilled upper Hubbard
bands by a modest value of on-site Coulomb repulsion. This
topologically non-trivial modification of the LDA electronic structure
due to the combined effects of spin-orbit coupling and on-site Coulomb
repulsion results in a Mott insulating state that is
antiferromagnetically ordered along $c$ axis with total moments
aligning antiparallel along the $c$ axis and canted along the $b$ axis. For
$U$ = 2.75 eV, the total magnetic moment is 0.67 $\mu_B$ with an
orbital contribution of 0.28 $\mu_B$ and a spin contribution of 0.38
$\mu_B$. These values differ from what is expected for the ideal
$J_\textrm{eff}$ = 1/2 state, and this deviation might be explained by
the mixing of $J_\textrm{eff}$ = 1/2 bands with Ir $e_g$ bands due to
the tilting of IrO$_6$ octahedra. There has been great interest in
finding different materials with unique magnetic properties. The results
presented here give strong support to the claim made by Ohgushi
\textit{et al.}\ in Ref.~\onlinecite{ohgu11} that CaIrO$_3$ has a
unique spin-orbit integrated magnetic ground state.

\section{Acknowledgements}
I am grateful to G.\ Jackeli, D.\ I.\ Khomskii, D.\ J.\ Singh, and I.\
I.\ Mazin for helpful discussions. I am thankful to L.\ Boeri for
encouragement, helpful discussions, and suggestions in improving the
manuscript, and to A. Avella for critical reading of the manuscript.


\begin{references}
\bibitem{foot:gap} In Ref.~\onlinecite{ohgu06}, the resistivity versus
    temperature curve is fit to the formula $\rho(T) = A \textrm{
      exp}(\Delta/T)$ and a value of $\Delta$ = 0.17 eV is
    obtained. The band gap should be twice the value of $\Delta$,
    i.e.\ 0.34 eV.
\end{references}
\end{document}